\documentclass[%
 reprint,
 amsmath,amssymb,
 aps,
]{revtex4-2}

\usepackage{graphicx}% Include figure files
\usepackage[font=small,labelfont=bf,
   justification=raggedright,singlelinecheck=0,
   format=plain]{caption}
   
\usepackage{subcaption}
\usepackage{physics}
\usepackage[table]{xcolor}
\usepackage{natbib,hyperref}
\hypersetup{
  colorlinks = true,
  linkbordercolor = {white},
  breaklinks=true,
  linkcolor = {black},
  anchorcolor = {black},
  citecolor = {black},
  filecolor = {black},
  menucolor = {black},
  runcolor = {black},
  urlcolor = {blue}
}

\begin{document}
\newcommand{\AJM}[1]{\textcolor{blue}{\textit{\textbf{Andrew:} #1}}}

\title{Finite Temperature Effects on the X-ray Absorption Spectra of Crystalline Aluminas from First Principles}

\author{Angela F Harper$^{1}$, Bartomeu Monserrat$^{1,2}$, \& Andrew J. Morris$^{3}$}

 \affiliation{%
    $^1$Theory of Condensed Matter, Cavendish Laboratory, University of Cambridge, J. J. Thomson Avenue, Cambridge CB3 0HE, U.K.\\%
    $^2$Department of Materials Science and Metallurgy, University of Cambridge, 27 Charles Babbage Road, Cambridge CB3 0FS, U.K.\\
    $^3$School of Metallurgy and Materials, University of Birmingham, Edgbaston, Birmingham B15 2TT, U.K.%
}
\email{a.j.morris.1@bham.ac.uk}

%IMAGE WORDCOUNT EST: 1266
%X-ray Absorption spectroscopy (XAS) is a high-resolution probe of the electronic density of states within solid-state materials. By including phonon-assisted transitions within plane-wave density-functional theory methods of calculating the XAS we obtain the Al K-edge XAS at 300\,K for two crystalline phases of alumina (Al$_2$O$_3$). From these finite-temperature spectra, we reproduce the pre-edge peak for the $\alpha$-Al$_2$O$_3$ phase, which arises from a mixed $s$-$p$ state that is not visible at the static-lattice level of approximation. In addition, we calculate the finite-temperature XAS for $\gamma$-Al$_2$O$_3$, which has not previously been reported, and find that the fully ordered spinel-based model of $\gamma$-Al$_2$O$_3$ describes two out of the three experimental transitions seen in the Al K-edge. We observe that the third peak in the calculated $\gamma$-Al$_2$O$_3$ spectrum arises from transitions in AlO$_6$ octahedral coordination environments from Al 1$s$ to delocalized $d$-like states. Finally, we propose this letter as a basis for future applications of finite-temperature XAS from first principles, as the methods used are fully generalizable to any atom and edge, beyond the Al K-edge. 

\begin{abstract}
By including phonon-assisted transitions within plane-wave DFT methods for calculating the X-ray absorption spectrum (XAS) we obtain the Al K-edge XAS at 300\,K for two Al$_2$O$_3$ phases. The 300\,K XAS reproduces the pre-edge peak for $\alpha$-Al$_2$O$_3$, which is not visible at the static-lattice level of approximation. The 300\,K XAS for $\gamma$-Al$_2$O$_3$ correctly describes two out of the three experimental peaks. We show that the second peak arises from 1s to mixed $s$-$p$ transitions and is absent in the 0\,K XAS. This letter serves as a basis for future applications, as the method is generalizable to any atom and edge.
\end{abstract}

\maketitle

Crystalline phases of alumina have a range of applications in ceramics and abrasives manufacturing \cite{medvedovski2006alumina,rao2014fabrication,paranjpe2017alpha}. 
Given the breadth of literature, especially in the case of corundum (the low energy $\alpha$-alumina phase), the crystal structure of $\alpha$-alumina may be deemed well-characterized by both experimental and computational methods.
Increasingly, as aluminas are used in electronic applications such as perovskite solar cells and heterogeneous catalysts \cite{trueba2005gamma,li2013post,das2020atomic}, it has become imperative to understand not only their atomistic but also their electronic structure, especially in the less well characterized phases such as $\gamma$-Al$_2$O$_3$. X-ray absorption spectroscopy (XAS) methods have been used to obtain the K-edge absorption spectra for $\alpha$-alumina \cite{kato2001quantification}, and more recently, $\gamma$-alumina \cite{altman2017chemical}, providing experimental insight into the electronic structure of these materials.

The Al K-edge XAS spectrum for $\alpha$-Al$_2$O$_3$ \cite{cabaret1996full} has a main absorption peak at $1568$\,eV, which accounts for transitions to 3$p$ states in the conduction band. Full multiple-scattering calculations reproduced this main peak, but failed to capture the pre-edge, predicted to arise from dipole-forbidden transitions from Al 1$s$ to 3$s$ states \cite{li1995k}. The pre-edge was later described using first-principles calculations which introduced the effect of atomic vibrations into the absorption cross section, with a method designed for cases in which the vibrational energies are small in relation to the absorption energy (as is the case for Al) \cite{Brouder2010Effect}. This confirmed that the existence of the pre-edge was allowed due to $s$-$p$ mixing arising from distortions to the local AlO$_6$ environment \cite{cabaret2009origin}. However this approximation assumes that the final electronic state of the system is not affected by vibrational modes and therefore calculates the excited electronic state without any vibrational effects included. Further, as is shown for the case of Ti \cite{Brouder2010Effect}, this method is unable to reproduce transitions between the Ti $s$ and $d$ states, as the assumption that the vibrational energies are small relative to the absorption energies no longer holds. While this approximation works for localized, small energy effect transitions, it is not generally applicable. 

Incorporating vibrational effects within first-principles calculations to obtain temperature dependent spectroscopy is the state-of-the-art.  For example, finite-temperature Nuclear Magnetic Resonance chemical shifts compare within 2-4\,ppm of experimental results for both organic \cite{monserrat2014temperature} and inorganic \cite{nemausat2015phonon} solid-state systems. Finite-temperature XAS of the Mg K-edge of MgO has also been studied, and a correlation between increasing temperature and decreasing absorption energy of the pre-edge peak was found in the Mg K-edge \cite{nemausat2015phonon}.  Despite these successes, finite temperature XAS has yet to be applied elsewhere, further underscoring the need for this study on the crystalline aluminas.

In this letter, we use first-principles XAS calculations combined with phonon calculations to produce finite-temperature Al K-edge XAS of both $\alpha$- and $\gamma$-alumina. By incorporating these finite-temperature effects at $300$\,K we are able to fully describe the pre-edge peak in $\alpha$-alumina, using a method for calculating phonon-assisted XAS, which accounts for vibrational effects across the whole unit cell and therefore represents a step change improvement over previous methods \cite{Brouder2010Effect,cabaret2009origin}. Additionally, we calculate the Al K-edge XAS spectra for $\gamma$-alumina, and assign both structural and electronic features to two of the three main edges in the $\gamma$-alumina XAS spectrum. We examine the electronic charge density of the third peak in the spectrum, and show that this peak is attributed to delocalized $d$-like states. This method of calculating finite-temperature XAS using first-principles accuracy can similarly be used on other crystalline structures, and provides an example of the utility of incorporating phonon effects in the modeling of XAS. 

X-ray Absorption Spectroscopy (XAS) measures the electronic transitions from a core state (in this case the Al $1s$ state for the Al K-edge spectrum) to the excited states in the conduction band of the material. The transitions are short range and vertical, and normally describe differences in energy states between nearest-neighbor atoms. When using pseudopotential plane-wave density-functional theory (DFT), which does not explicitly treat electrons in the core orbitals, it is necessary to approximate the effect of the X-ray excitation of an electron by incorporating a pseudopotential at the site of the excitation which has the electronic configuration of the atom with an electron removed from the core state \cite{gao2009core}. This core-hole pseudopotential method has shown success in reproducing the experimental absorption spectra of a range of materials  \cite{mizoguchi2009first,yamamoto2004first,PhysRevB.75.184205} as implemented in the DFT code, CASTEP \cite{clark2005first}.  To recover the wave function of the all electronic core state requires a transformation using a projector-augmented wave approach in which each transmission matrix element is calculated from a linear transformation of the pseudo wavefunction \cite{gao2009core}. 

The X-ray absorption cross section, $\sigma$, is described using Fermi's golden rule approximation to the imaginary part of the dielectric function \cite{nemausat2015phonon,fermi1950nuclear},

\begin{equation} \label{eq:dipole}
\sigma(\omega)= 4 \pi^{2} \alpha_{0} \hbar \omega \sum_{f}\left|\left\langle\Psi_{f}|\hat{\boldsymbol{\epsilon}} \cdot \mathbf{r}| \Psi_{i}\right\rangle\right|^{2} \delta(E_{f}-E_{i}-\hbar \omega),
\end{equation}
where $\hbar\omega$ is the energy of the incoming X-ray, $\mathbf{r}$ is the single electron position operator, $\alpha_{0}$ is the fine structure constant, and $\hat{\boldsymbol{\epsilon}}$ is the polarization direction of the electromagnetic
vector potential. The incoming X-ray excites an electron from a core-level orbital $|\Psi_i\rangle$ to a final state $|\Psi_{f}\rangle$ in the conduction band. The final electronic states, $|\Psi_{f}\rangle$, are eigenstates of the Kohn-Sham Hamiltonian.

The zero temperature approximation of the X-ray absorption cross section (Equation \ref{eq:dipole}) can be extended to finite temperature using the Williams-Lax theory \cite{williams,lax1952franck,monserrat2018electron,nemausat2015phonon}, which yields the following expression,

\begin{equation} \label{eq:dipole_T}
\begin{split}
\mathcal{\sigma}(T)=& \frac{1}{\mathcal{Z}} \sum_{k}\left\langle\chi_{k}(\mathbf{R})\right| \mathcal{\sigma}(\omega,\mathbf{R}) \left|\chi_{k}(\mathbf{R})\right\rangle \mathrm{e}^{-E_{k} / k_{\mathrm{B}} T},
\end{split}
\end{equation}
in which $\mathcal{Z}=\sum_{k} \mathrm{e}^{-E_{k} / k_{\mathrm{B}} T}$ is the partition function from the initial state, where $\mathrm{E_{k}}$ is the energy of vibrational state $k$, and $\ket{\chi_{k}(\mathbf{R})}$ is the vibrational wave function with nuclear configuration $\mathbf{R}$, which in this letter we describe within the harmonic approximation. Using a Monte Carlo (MC) sampling technique \cite{monserrat2018electron} we evaluate Equation \ref{eq:dipole_T}, by calculating a series of X-ray absorption transition energies, for sampled nuclear configurations $\textbf{R}$ distributed according to the harmonic vibrational density.

The finite temperature XAS energies at $300$\,K were calculated for both $\alpha$- and $\gamma$-Al$_2$O$_3$, with a phonon \textbf{q} point grid of 2$\times$2$\times$2, and 30 MC sampling points. The ground-state, static-lattice unit cell of  $\alpha$-Al$_2$O$_3$, has one symmetry equivalent Al site, in an octahedral AlO$_6$ environment, so that a core-hole was placed on one site in each of the 30 MC configurations to calculate the XAS at 300\,K. 

\begin{figure}[htb!]
  \centering
  \includegraphics[width=0.48\textwidth]{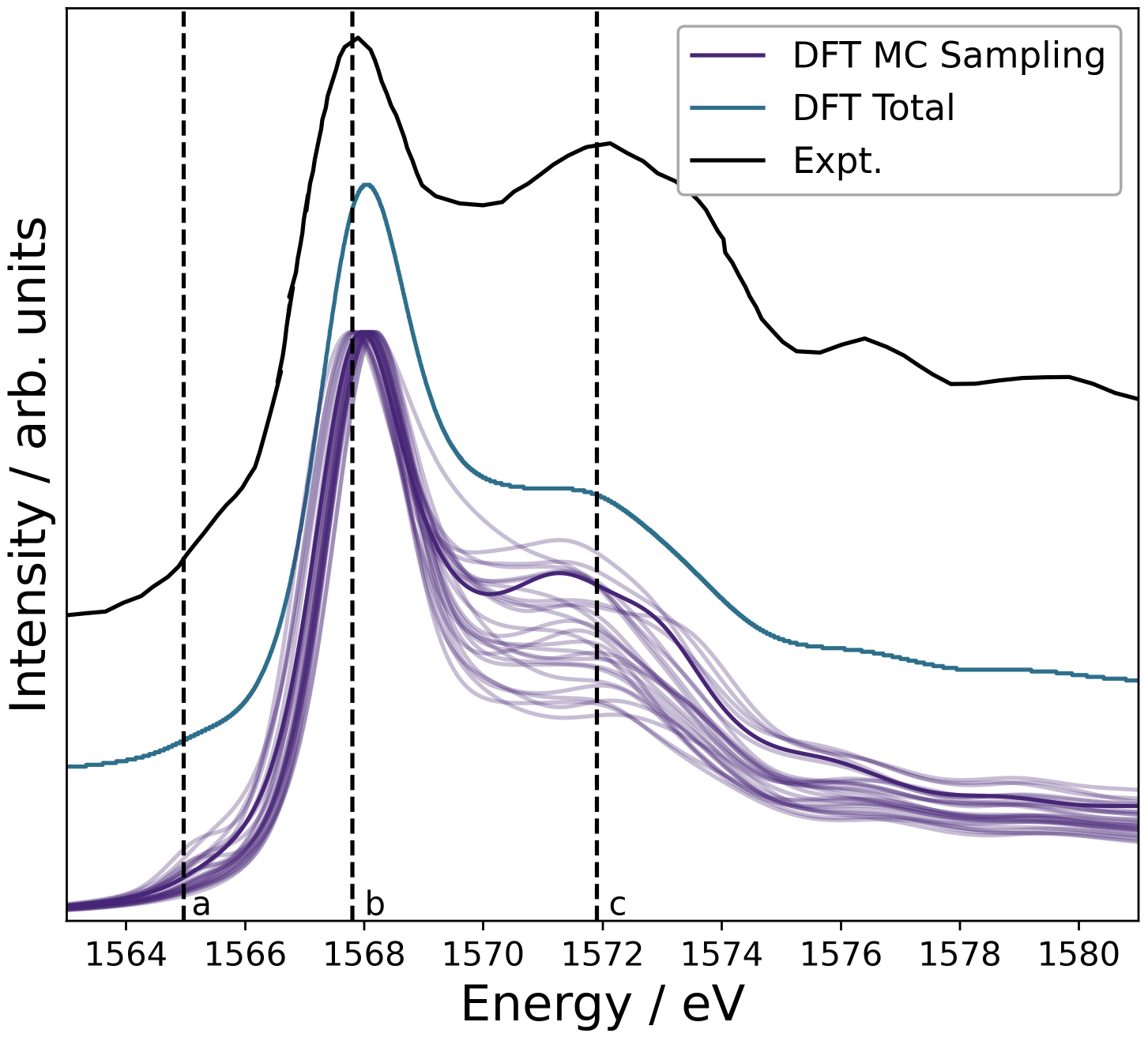}
  \caption{Finite temperature XAS spectrum (blue) calculated as a sum of 30 spectra from the MC sampled configurations of $\alpha$-{Al2O3} (purple), compared to the experimental XAS spectra from Cabaret and Brouder, \textit{Journal Physics Conference Series}, 2009 \cite{cabaret2009origin} indicating that the pre-peak at \textbf{a} is visible in the MC sampled spectra.}.
  \label{fig:alpha-xas}
\end{figure}

The experimental Al K-edge absorption spectrum for $\alpha$-Al$_2$O$_3$ contains three peaks (\textbf{a}, \textbf{b}, and \textbf{c}) at $1565$\,eV, $1568$\,eV, and $1572$\,eV respectively, shown in Figure \ref{fig:alpha-xas}.  The pre-edge peak at \textbf{a} is assigned to transitions from Al 1$s$ to 3$s$ states \cite{altman2017chemical}. At $1568$\,eV, peak \textbf{b} is the main absorption peak in the Al octahedral sites and results from a transition from the Al 1$s$ to 3$p$ state. Peak \textbf{c} is attributed to transitions from the Al 1$s$ to 3$d$ states \cite{altman2017chemical}.

%292 words 0.55 ratio

%359 words 1.88461538462 ratio
\begin{figure}[htb!]
    \centering
    \includegraphics[width=0.48\textwidth]{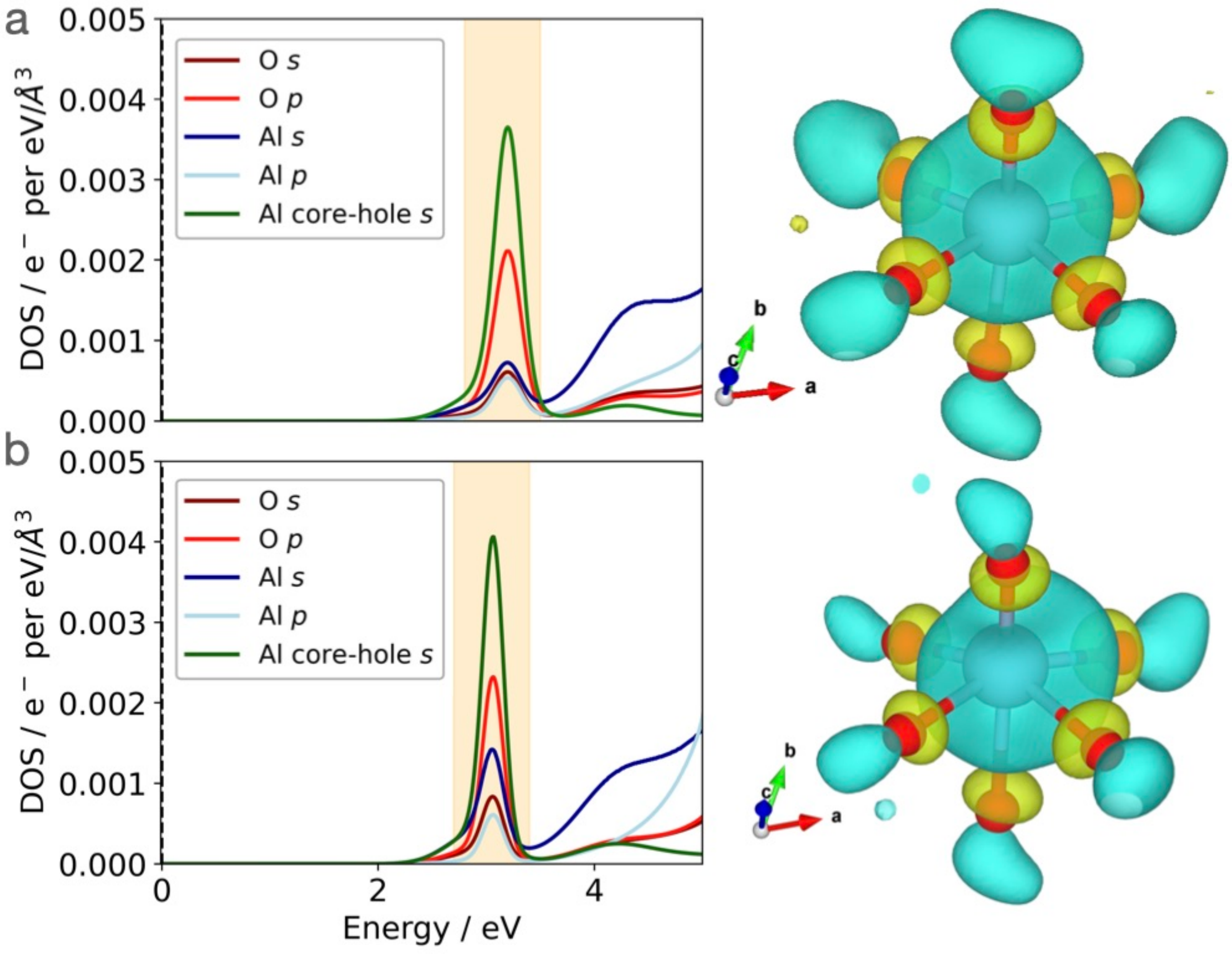}
    \caption{Electronic DOS for the ground-state static lattice structure of $\alpha$-Al$_2$O$_3$ and visualization of the states at 3\,eV surrounding the core-hole Al atom. (b) Electronic DOS and visualization for one MC sampling configuration of $\alpha$-Al$_2$O$_3$.  The orange shaded region in the DOS is the energy at which the electron density is calculated. In the right panels, Al atoms are shown in blue, O atoms in red, and the positive and negative electron density in 3D.}
    \label{fig:alpha-dos}
\end{figure}

 The ground-state static-lattice XAS calculation of $\alpha$-Al$_2$O$_3$ reproduces peaks \textbf{b} and \textbf{c} (Figure \ref{fig:gamma-static-xas}a), but the pre-edge at \textbf{a} is absent. This pre-edge comprises an $s$-to-$s$ transition, hence is dipole forbidden due to the inversion symmetry present in the unit cell. However, the pre-edge peak at \textbf{a}, as well as peaks \textbf{b} and \textbf{c} are visible in the XAS calculated at 300\,K, indicating that the inclusion of phonon assisted transitions successfully describes all three peaks in the $\alpha$-Al$_2$O$_3$ Al K-edge XAS. 
 
 By calculating the electronic density of states (DOS) for both the ground-state static lattice case, and one configuration from the MC sampling, we can determine the orbital character of the states at this pre-edge. In both the ground-state static lattice and MC sampling configuration DOS, there are Al $3s$ states at 3\,eV above the Fermi level, however in the MC sampling case there are additional O $p$ states at this energy level.  The presence of the O $p$ states is a result of the distortion of the lattice at 300\,K breaking the inversion symmetry in the unit cell. This additionally distorts the local octahedral environment surrounding the Al atom (shown in Figure \ref{fig:alpha-dos}). Comparing the static lattice and MC sampled sites based on their continuous symmetry measure (CSM) as described by Pinsky \cite{pinsky1998continuous}, gives a quantitative measure of the site's distortion from the pristine octahedral symmetry. Higher CSM values show further deviation; the AlO$_6$ site in the static lattice unit cell has a symmetric energy density around the Al atom (CSM = 0.59), and in the MC sampled configuration this site has lower symmetry (CSM = 0.81) \cite{waroquiers2017statistical}. Thus, while the Al 3$s$ states are present in both cases, the transitions at the pre-edge are allowed only in the MC sampled configuration, due to the presence of a mixing of $s$ and $p$ states and breaking of inversion symmetry.

%156 words 1.10426540284 ratio
\begin{figure}
    \centering
    \includegraphics[width=0.48\textwidth]{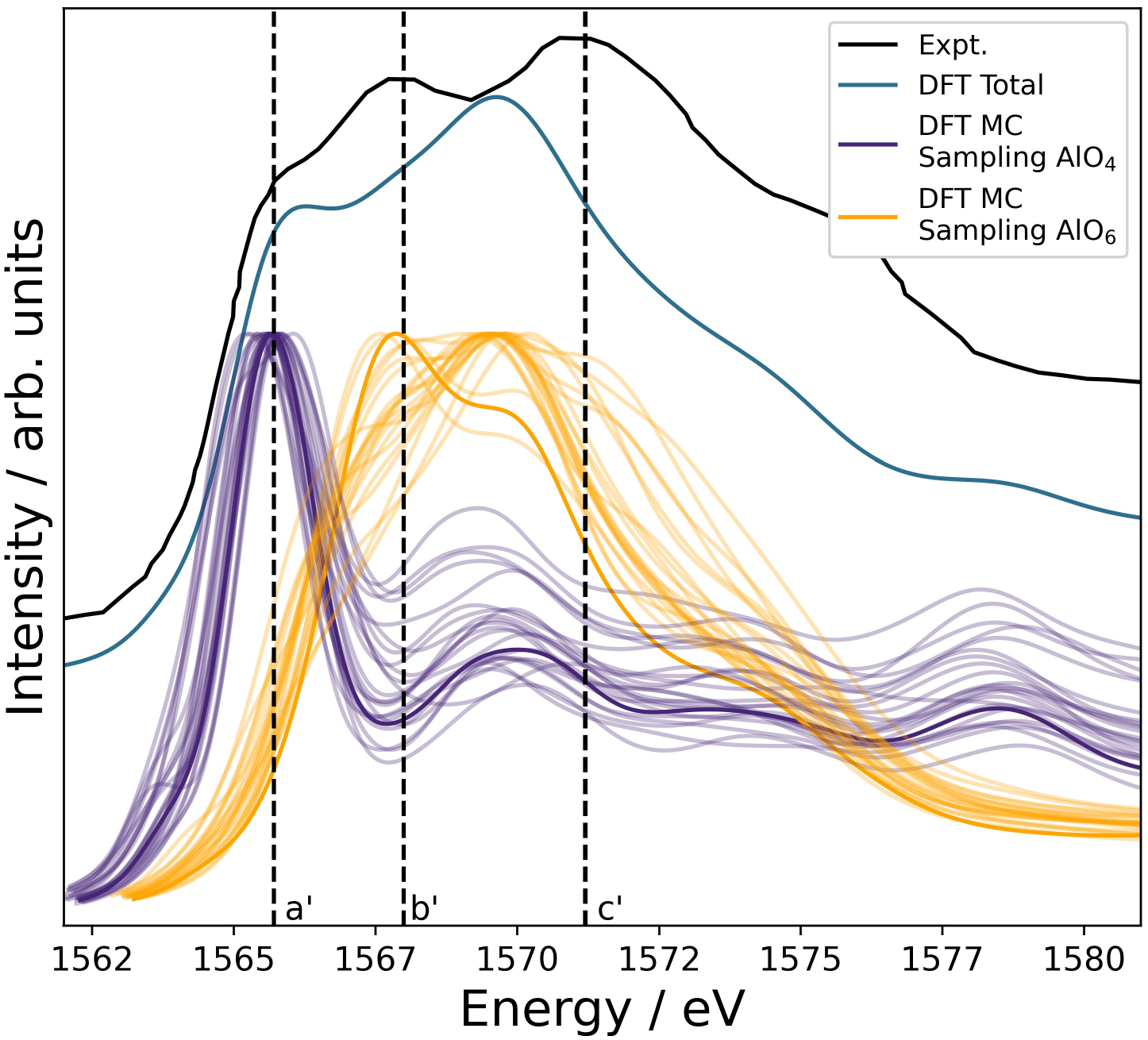}
    \caption{Finite temperature XAS for $\gamma$-Al$_2$O$_3$ calculated as the weighted average over 30 generated configurations from Monte Carlo sampling over the static lattice phonon modes. Peak \textbf{a$'$} is well described the spectra with a core-hole on the AlO$_4$ site. Peak \textbf{b$'$} is described by six spectra with a core-hole on the AlO$_6$ site, and the remaining spectra with a core-hole on the AlO$_6$ site have a peak at $1570.0$\.eV. Only the \textbf{c$'$} peak from experiment is not described by the finite temperature spectrum. Experimental spectrum adapted with permission from Altman \textit{et al.}. \textit{Inorg. Chem.} 2017, 56, 10, 5710–5719 Copyright 2017 American Chemical Society.}
    \label{fig:gamma-xas}
\end{figure}

%471 words?? ratio 1.38920134983
\begin{figure*}[!htb]
    \centering
    \includegraphics[scale=0.35]{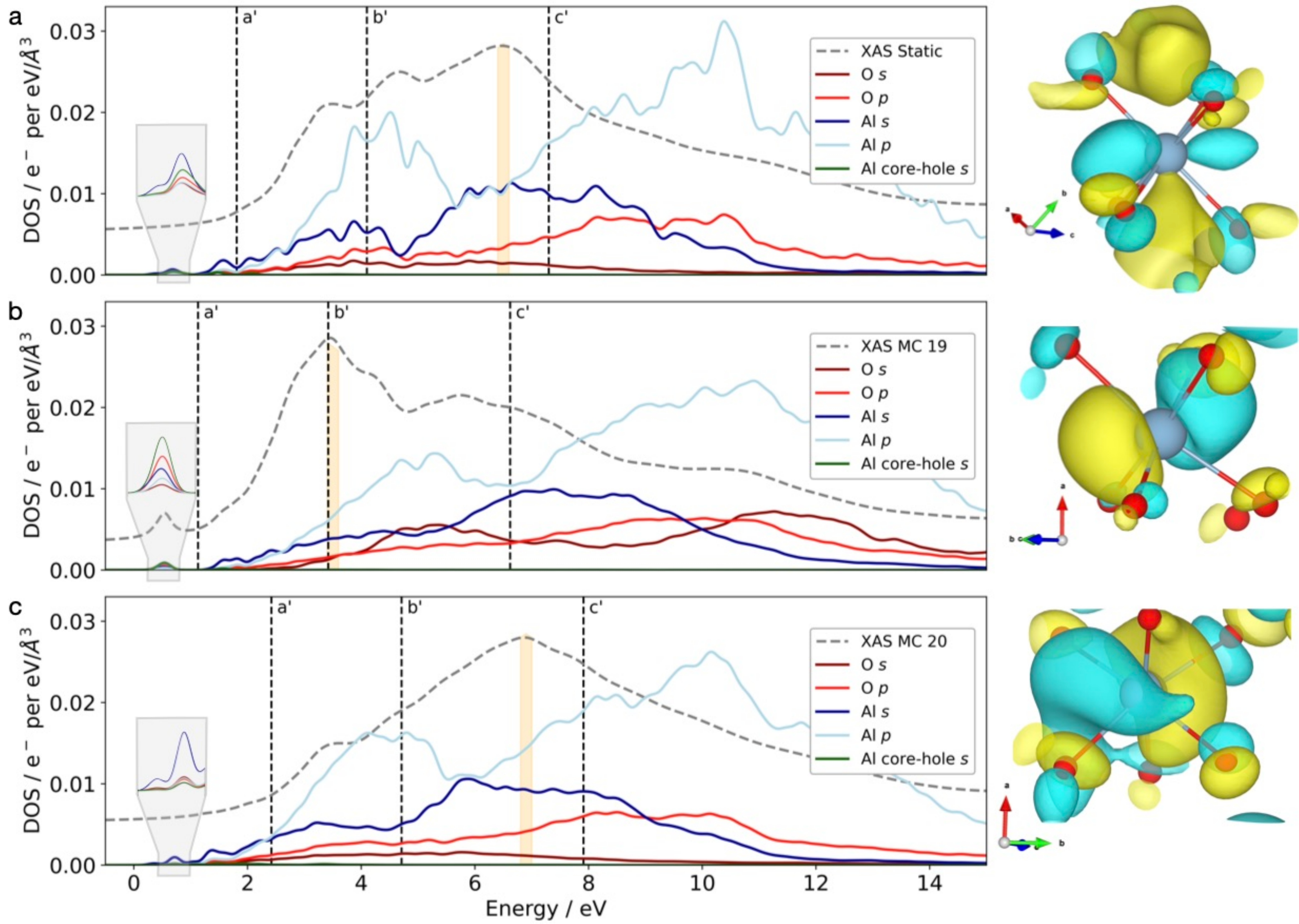}
    \caption{Electronic DOS for the static-lattice structure of $\gamma$-Al$_2$O$_3$ and two configurations from MC sampling at 300\,K alongside the orbital character of the states giving rise to the main peak in the XAS for each case. The Fermi level is at 0 eV, and the XAS is aligned to the corresponding energy levels in the DOS. The orange shaded region in the DOS is the point at which the electron density is calculated. The gray shaded region highlights the Al core-hole $s$ states near the CBM. In the right panels, Al atoms are shown in blue, O atoms in red, and the positive and negative electron density in 3D. (a) Electronic DOS for the static lattice structure of $\gamma$-Al$_2$O$_3$ and visualization of the Al \textit{d}-like states giving rise to the maximum between \textbf{b$'$} and \textbf{c$'$} in the XAS. (b) Electronic DOS for MC sampling configuration 19, and visualization of mixed $s$-$p$ states corresponding to a maximum peak in the XAS of $1568.0$\,eV at \textbf{b$'$}. (c) Electronic DOS for MC sampling configuration 20, and visualization of delocalized electronic orbitals which cannot easily be assigned to $s$, $p$, or $d$ character, corresponding to an XAS peak at $1570.0$\,eV between \textbf{b$'$} and \textbf{c$'$}.}
    \label{fig:gamma-dos}
\end{figure*}

The Al K-edge XAS spectrum for $\gamma$-alumina is less well-understood than that of $\alpha$-Al$_2$O$_3$, despite the large number of studies on the K-edge spectrum of aluminosilicate glasses \cite{de2004and,neuville2004environment,neuville2004speciation}. Unlike $\alpha$-alumina, which contains only one, symmetry-equivalent, octahedral AlO$_6$ site, in the $\gamma$-Al$_2$O$_3$ phase, there are two symmetry-inequivalent sites, an octahedral AlO$_6$ site and tetrahedral AlO$_4$ site. The cation disorder in this system has lead to several model unit cells of $\gamma$-alumina \cite{ayoola2020evaluating}. Discrepancies between experimental X-ray diffraction and first principles models suggest a variety of potential models for $\gamma$-Al$_2$O$_3$: namely a spinel model in which the Al$^{3+}$ cations are all on spinel sites \cite{smrvcok2006gamma}; a spinel model in which these cations sit on non-spinel sites \cite{PintoGammaAl2O3}; and a non-spinel model \cite{digne2004use,paglia2003tetragonal}. In this work, we have chosen to use the Pinto \textit{et al.} phase of $\gamma$-Al$_2$O$_3$ \cite{PintoGammaAl2O3}, as recent selected-area electron diffraction analysis suggests that the Pinto \textit{et al.} model shows excellent agreement with experiment \cite{ayoola2020evaluating,prins2020structure}. 

The presence of two symmetry-inequivalent sites in $\gamma$-Al$_2$O$_3$ requires a careful treatment of the contribution of these two sites to the overall XAS lineshape. Typically, total XAS spectra are weighted according to the abundance of each symmetry-inequivalent atom in the system \cite{klein2021nuts}; thus to calculate the $\gamma$-Al$_2$O$_3$ XAS, we weight the contribution of core-holes placed on AlO$_4$ and AlO$_6$ sites according to their abundance in the static lattice unit cell (10 AlO$_6$ to 6 AlO$_4$ units).

The experimental Al K-edge XAS of $\gamma$-Al$_2$O$_3$ contains an edge (\textbf{a$'$}) at $1565.7\,$eV, and two peaks (\textbf{b$'$} and \textbf{c$'$}) at $1568.0$ and $1571.2$\,eV respectively. Edge \textbf{a$'$} and peak \textbf{b$'$} have been assigned to $1s$ to $3p$ transitions and \textbf{c$'$} is assigned to $1s$ to $3d$ transitions \cite{altman2017chemical}. Peak \textbf{c$'$} at $1571.2$\,eV is also present in $\alpha$-Al$_2$O$_3$ XAS and other aluminosilicate materials, corresponding to transitions to Al $3d$ states \cite{altman2017chemical}. The ground-state static lattice XAS with a core-hole placed on the AlO$_4$ site reproduces the peak at \textbf{a$'$}, but the XAS with a core-hole placed on the AlO$_6$ site has a peak with an absorption energy between \textbf{b$'$} and \textbf{c$'$} at $1570.0$\,eV. The static lattice XAS (Figure \ref{fig:gamma-static-xas}b) describes neither peak \textbf{b$'$}or \textbf{c$'$}, and therefore justifies the use of the finite temperature XAS at $300$\,K, in order to resolve the unassigned peaks.

At 300K, the first principles XAS for $\gamma$-Al$_2$O$_3$ has peaks at 1565.7\,eV, 1568.0\,eV, and 1570.0\,eV, as shown in Figure \ref{fig:gamma-xas}. The peak at 1565.7\,eV corresponds to \textbf{a$'$} in experiment. The finite temperature XAS successfully reproduces peak \textbf{b$'$} in a set of 6 MC sampled configurations with a core-hole on the AlO$_6$ site. This peak is not described by the static lattice XAS for $\gamma$-Al$_2$O$_3$ shown in Figure \ref{fig:gamma-static-xas}, highlighting the need for incorporating temperature effects. Finally, the third peak in the XAS at 300\,K is 1.2\,eV lower than the \textbf{c$'$} transition, and is present in a set of 24 MC configurations with a core-hole placed on the AlO$_6$ site.

By calculating the electronic DOS for the static lattice $\gamma$-Al$_2$O$_3$ structure and two configurations from MC sampling, we can determine the orbital character of the states which give rise to peak \textbf{b$'$} at 1568.0\,eV and the peak at  $1570.0$\,eV. In the ground-state static lattice unit cell with a core-hole placed on the AlO$_6$ site, shown in Figure \ref{fig:gamma-dos}a, the electron density at the $1570.0$\,eV peak in the XAS has $d$-like character ($6.5$\,eV above the Fermi level in the DOS). 

The electronic DOS for one MC sampling configuration shown in Figure \ref{fig:gamma-dos}b, with a peak at \textbf{b$'$}, shows the states at \textbf{b$'$} have a mixed $s$-$p$ character. This peak has previously been assigned to Al 1$s$ to 3$p$ states \cite{altman2017chemical}, however these results suggest that the orbital character of these states is mixed $s$-$p$. Figure \ref{fig:gamma-dos}c also shows the DOS for a configuration from MC sampling which has a peak at $1570.0$\,eV in the XAS. The states which contribute to the peak at 1570.0\,eV are delocalized, and cannot be easily assigned to an $s,p$ or $d$ character.

Peak \textbf{c$'$} in the XAS in Figure \ref{fig:gamma-xas} is not described by any of the spectra with core-holes placed on the AlO$_6$ site in the MC configurations at 300\,K. However, the $d$-like states in the static lattice spectrum in Figure \ref{fig:gamma-dos}a, and delocalized states in the configuration from MC sampling in \ref{fig:gamma-dos}c, are suggestive of transitions to $d$-like states as in experiment at peak \textbf{c$'$}. Although the DFT transition energy of this third peak is at a lower energy than in experiment, this is a good first approximation for visualizing these mixed states at peak \textbf{c$'$}. One potential source of this discrepancy is in the use of the PBE functional, and could be resolved through an investigation of the electronic states in the region above the Fermi level using a hybrid functional. While this is beyond the scope of this work, which focuses on the finite temperature DOS, this could resolve the energy difference between \textbf{c$'$} experiment and theory. Using a multiple-scattering approach to calculating the XAS would resolve features at higher energies, however our work here determines which transitions are a function of thermal vibrations and obtain accurate transition energies near the absorption edge using DFT. However, given the strong agreement between theory and experiment for peaks \textbf{a$'$} and \textbf{b$'$} there is reason to suggest that the structure is the source of this energy difference. By calculating the XAS of another $\gamma$-Al$_2$O$_3$ structure, one could explore the effects of partial occupancy on the electronic configurations.

The electronic structure of both $\alpha$- and $\gamma$-Al$_2$O$_3$ has been characterized experimentally using Al K-edge XAS, in order to describe the electronic transitions present in these two crystalline phases of Al$_2$O$_3$ \cite{cabaret1996full,cabaret2009origin,altman2017chemical}. Using MC sampling on the vibrational modes of the ground state structure of these alumina phases has shed light onto the origins of their electronic transitions in the XAS, and suggested possible further studies for $\gamma$-Al$_2$O$_3$. The finite temperature XAS for $\alpha$-Al$_2$O$_3$ accurately reproduces the absorption pre-edge not observed at the static lattice level of theory and the resulting electronic states above the Fermi level show $s$-$p$ mixing. The ground state XAS for $\gamma$-Al$_2$O$_3$ only describes the K-edge transitions in the AlO$_4$ sites. By incorporating finite temperature effects, we are able reproduce two of three peaks in the experimental Al K-edge XAS. 

Beyond the results we have shown for the crystalline phases of $\alpha$- and $\gamma$-Al$_2$O$_3$, we have demonstrated that using the Williams-Lax theory together with a stochastic configuration sampling technique \cite{williams,lax1952franck,monserrat2018electron} is a general method for calculating finite-temperature XAS. By calculating the corresponding DOS of states for each spectrum we gain additional information which is not accessible in experiment, including visualizing the orbitals alongside their absorption peaks. Thus the XAS calculated at 300\,K for $\alpha$- and $\gamma$-Al$_2$O$_3$ not only contains the same peak positions, but also the same assigned orbital transitions as experiment. This method is not specific to crystalline aluminas, the K-edge absorption spectrum, or to X-ray Absorption spectroscopy, and can be easily applied to other systems and absorption edges in order to study experimental features not observed at the ground-state static-lattice level.

\section{Acknowledgments}
 AFH would like to thank Mike Payne for his support and helpful discussions. AFH acknowledges the financial support of the Gates Cambridge Trust and the Winton Programme for the Physics of Sustainability, University of Cambridge, UK. BM acknowledges support from a UKRI Future Leaders Fellowship (MR/V023926/1), from the Gianna Angelopoulos Programme for Science, Technology, and Innovation, and from the Winton Programme for the Physics of Sustainability. AJM acknowledges funding from EPSRC (EP/P003532/1). The authors acknowledge networking support via the EPSRC Collaborative Computational Projects, CCP9 (EP/M022595/1) and CCP-NC (EP/T026642/1). The calculations in this letter were performed using resources provided by the Cambridge Service for Data Driven Discovery (CSD3) operated by the University of Cambridge Research Computing Service (www.csd3.cam.ac.uk), provided by Dell EMC and Intel using Tier-2 funding from the Engineering and Physical Sciences Research Council (capital grant EP/P020259/1), and DiRAC funding from the Science and Technology Facilities Council (www.dirac.ac.uk).

% \section{Supporting Information Description}

% The supporting information contains four figures cited in the main text which contribute to the understanding of both the methods and results of this work, and the figure captions describe their relevance to the work in the main text. All of the calculations and data for this paper will be made available on the University of Cambridge's Data Repository Apollo at \href{https://www.repository.cam.ac.uk/}{https://www.repository.cam.ac.uk/}.

\bibliography{apssamp}

\clearpage
\begin{center}
\textbf{\large Supplemental Materials}
\end{center}
\captionsetup[subfigure]{labelfont=bf,textfont=normalfont,font=small,singlelinecheck=off,justification=centering}
\setcounter{equation}{0}
\setcounter{figure}{0}
\setcounter{table}{0}
\setcounter{page}{1}
\makeatletter
\renewcommand{\theequation}{S\arabic{equation}}
\renewcommand{\thefigure}{S\arabic{figure}}

\subsection{Dynamical Stability}
\begin{figure*}[htb!]
    \centering
    \begin{subfigure}{.6\textwidth}
    \includegraphics[width=\linewidth]{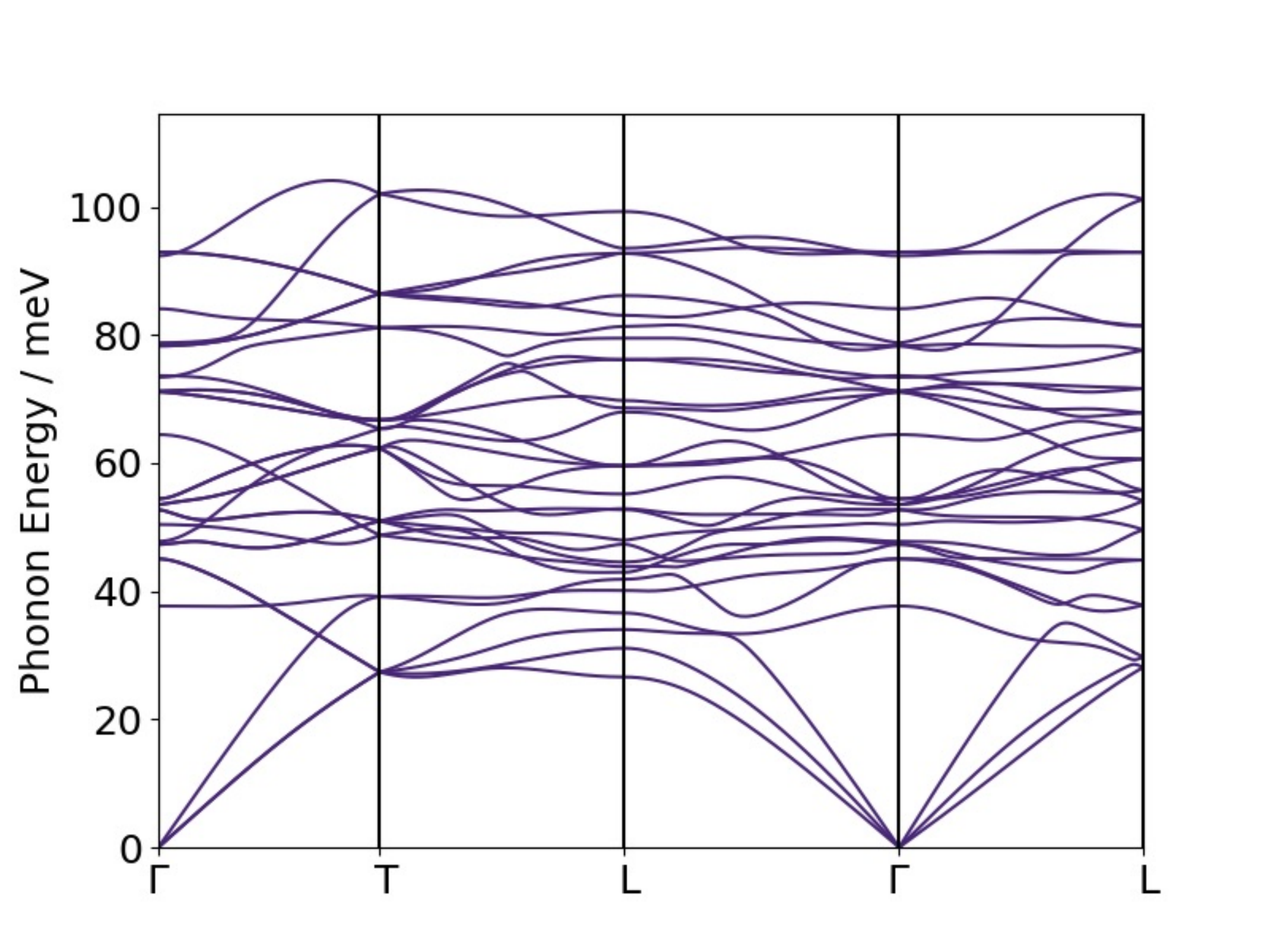}
    \caption{}
    \end{subfigure}
    \begin{subfigure}{.6\textwidth}
    \includegraphics[width=\linewidth]{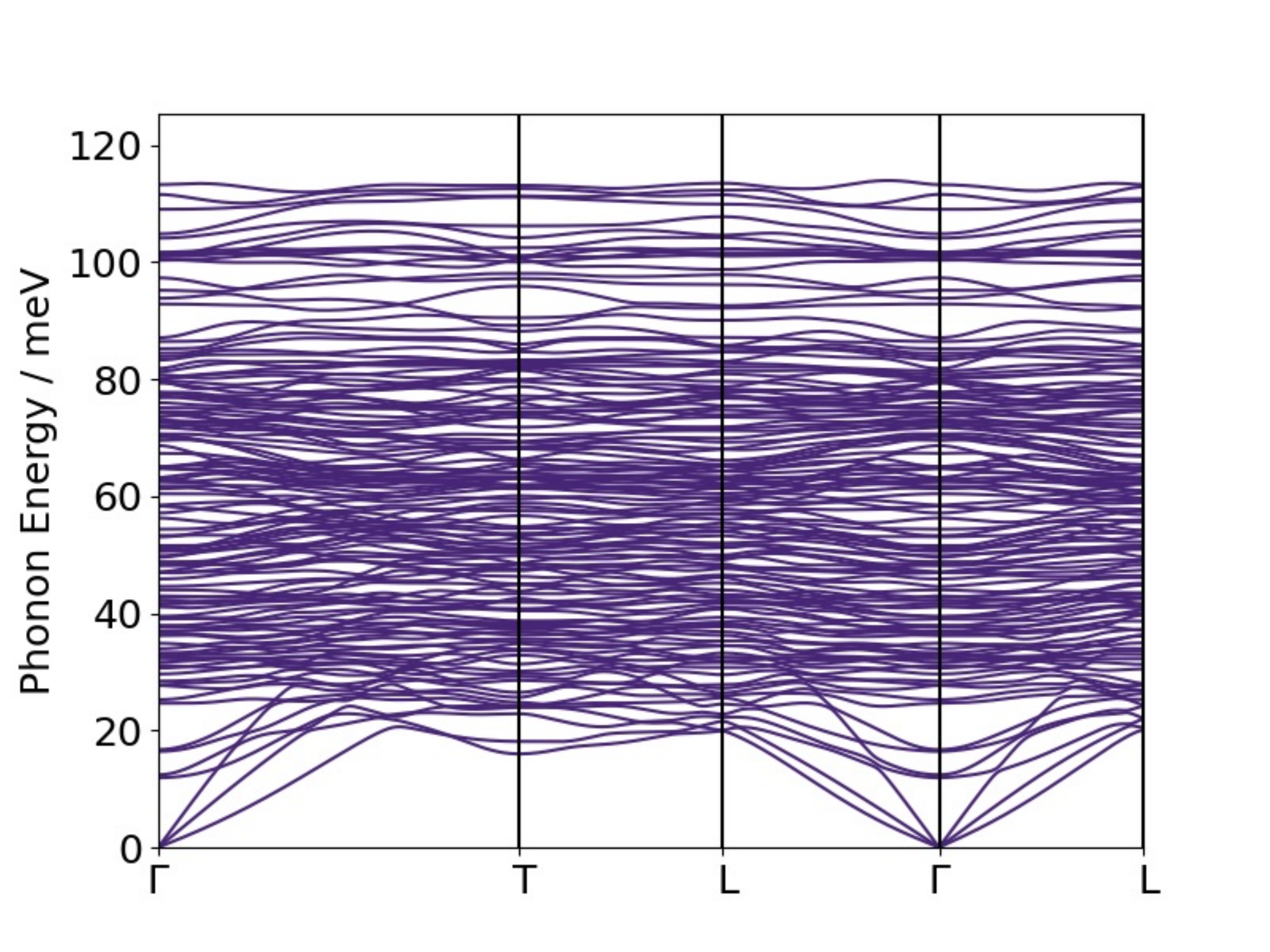}
    \caption{}
    \end{subfigure}
    \caption{(a) $\alpha$-Al$_2$O$_3$ phonon dispersion through high symmetry points $\Gamma$-$T$-$S$-$\Gamma$-$L$ (b) $\gamma$-Al$_2$O$_3$ phonon dispersion through high symmetry points $\Gamma$-$T$-$S$-$\Gamma$-$L$. Both using (a) and (b) at a plane-wave energy cut-off of $700$\,eV using the PBE functional at a \textbf{q}-point grid size of 2x2x2. This phonon dispersion is used to construct the supercells from Monte Carlo sampling, which were then used to calculate the XAS spectra at finite temperature. }
    \label{fig:y-phonons}
\end{figure*}

The atomic positions in both $\alpha$- and $\gamma$-Al$_2$O$_3$ were optimized to reduce the forces to within $0.01$\,eV/\r{A} at a plane-wave basis set cut off of $700$\,eV and an electron \textit{k}-point grid size of 4$\times$4$\times$4.  In Figure \ref{fig:y-phonons} we show the phonon dispersion for both $\alpha$- and $\gamma$-Al$_2$O$_3$ calculated with a coarse \textbf{q}-point grid of 2$\times$2$\times$2. Using finite differences combined with the nondiagonal supercell method of constructing commensurate supercells \cite{PhysRevB.92.184301}, a total of 4 supercells were required for $\alpha$-Al$_2$O$_3$, which has 10 atoms in the unit cell. For $\gamma$-Al$_2$O$_3$, which contains 40 atoms in the unit cell, a total of 6 supercells were required to calculate the phonon dispersion. The number of supercells required is based on the symmetry of the unit cell \cite{PhysRevB.92.184301}. There are no imaginary phonon modes at this level of theory, as shown in Figure \ref{fig:y-phonons}, indicating that both structures are dynamically stable.

\begin{figure*}[hbt!]
    \centering
    \includegraphics[width=0.8\textwidth]{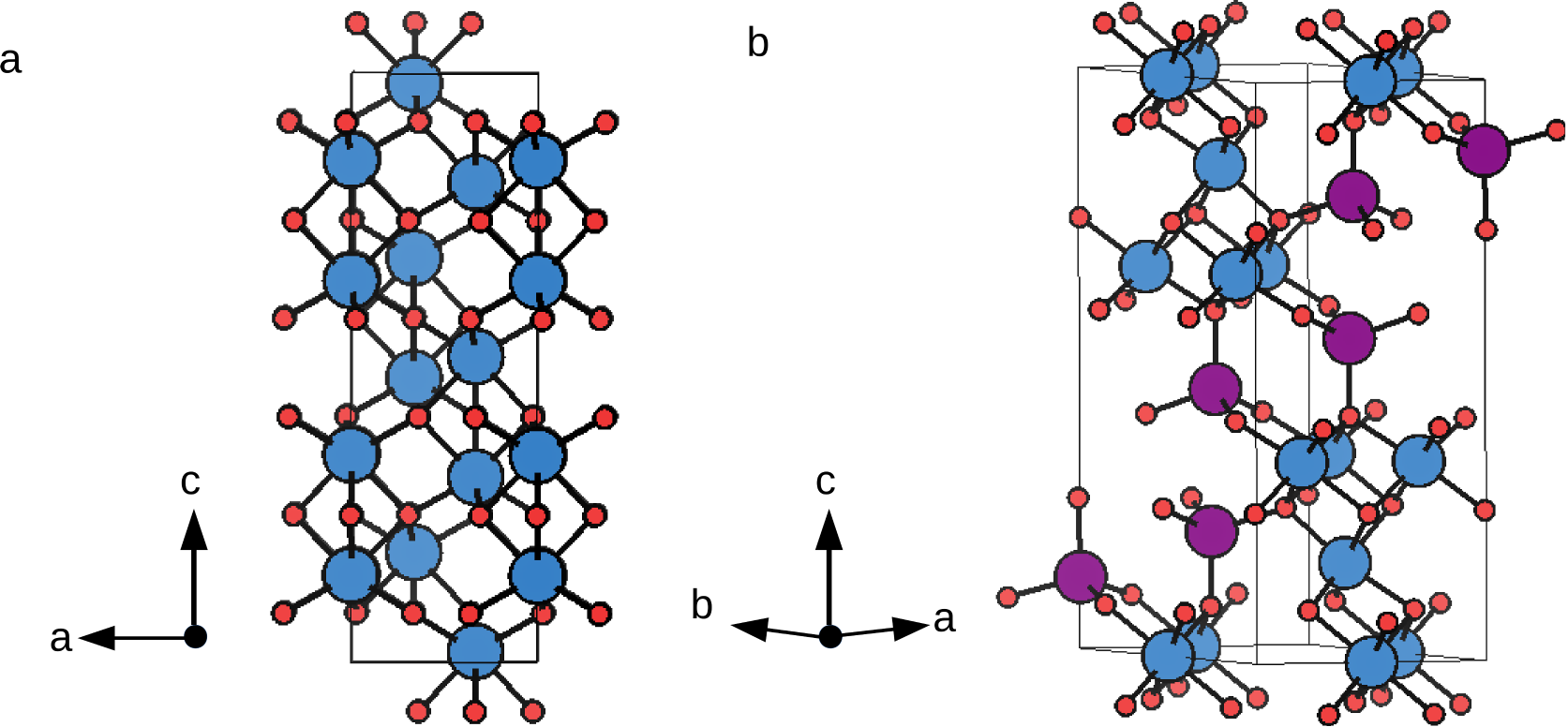}
    \caption{(a) The $\alpha$-Al$_2$O$_3$ structure has $R\bar{3}c$ space group with one symmetry-equivalent Al environment. All Al atoms are in an octahedral arrangement (blue) and all O atoms are in tetrahedral environments. All O atoms are shown in red, with Al-O bonds shown in black. (b) Pinto \textit{et al.} fully-ordered spinel structure of $\gamma$-Al$_2$O$_3$ \cite{PintoGammaAl2O3}, containing two Al vacancies on the octahedral site of the $C2/m$ unit cell. This phase contains both octahedral AlO$_6$ environments (blue) and tetrahedral AlO$_4$ environments (purple).}
    \label{fig:structures}
\end{figure*}

The fully ordered Pinto \textit{et al.} structure of $\gamma$-Al$_2$O$_3$ is used here in the calculation of the XAS spectrum of $\gamma$-Al$_2$O$_3$ and is shown alongside the $R\bar{3}m$ phase of $\alpha$-Al$_2$O$_3$ in Figure \ref{fig:structures}. Other phases which have partial occupancy on the AlO$_6$ site would require additional analysis of the site orderings, and the use of large supercell models with 300 atoms, which are too computationally intensive for first principles XAS. Ayoola \textit{et al.} \cite{ayoola2020evaluating} suggest that of the postulated structures of $\gamma$-Al$_2$O$_3$ in the literature, the Smr\v{c}ok cubic spinel model could represent disorder that is not present in the fully ordered Pinto model used in this work. However, the use of the Pinto model is partly justified as it contains the same ratio of tetragonal to octahedral Al environments as the Smr\v{c}ok model  \cite{PintoGammaAl2O3,ayoola2020evaluating}.

\subsection{Transition Energy Referencing}

The spectral features for an XAS calculation are produced using the method above, however a correction is needed, as described by Mizoguchi \textit{et al.} \cite{mizoguchi2009first}, to calculate the theoretical transition energy ($E_\mathrm{TE}$). In an all-electron calculation, $E_\mathrm{TE}$ is directly computed from the total energy difference between the excited and ground state electron configurations, but when using pseudopotentials only the valence electrons are accounted for in the total energy. Thus, calculating the transition energy requires a contribution from both core and valence electron energies, such that

\begin{equation}
    E_\mathrm{TE} = \Delta E_{\mathrm{valence}} + \Delta E_{\mathrm{core(atom)}}
    \label{eq:e_te}
\end{equation}
where

\begin{equation}
    \Delta E_{\mathrm{core(atom)}} = \Delta E_{\mathrm{All\,orbitals(atom)}} - \Delta E_{\mathrm{valence(atom)}}.
    \label{eq:core_atom}
\end{equation}
In Equation \ref{eq:e_te}, $E_{\mathrm{TE}}$ is the total transition energy, calculated by summing the energy difference in the valence electron density, $\Delta E_{\mathrm{valence}}$, and in the core orbital energy for the core-hole pseudopotential, $\Delta E_{\mathrm{core(atom)}}$ between the core-hole and non-core-hole calculations. The $\Delta E_{\mathrm{core(atom)}}$ is the difference between the all-electron energy and pseudopotential energy for the atom with the core-hole, $\Delta E_{\mathrm{All\,orbitals(atom)}}$, and the all-electron versus pseudopotential energy for the valence electrons, $\Delta E_{\mathrm{valence(atom)}}$ from the core-hole pseudopotential.

Equation \ref{eq:core_atom} calculates the transmission energy from the core-hole pseudopotential calculation $\Delta E_{\mathrm{core(atom)}}$, whereas $E_{\mathrm{TE}}$ and $\Delta E_{\mathrm{valence}}$ are obtained from the full pseudopotential XAS calculation. By comparing with experimental results it is possible to calculate both the total energy shift as well as the relative shifts between core-hole calculations for several XAS calculations.

\subsection{0 K Static Lattice XAS}

\begin{figure*}[htb!]
    \centering
    \begin{subfigure}{.5\textwidth}
    \includegraphics[width=\linewidth]{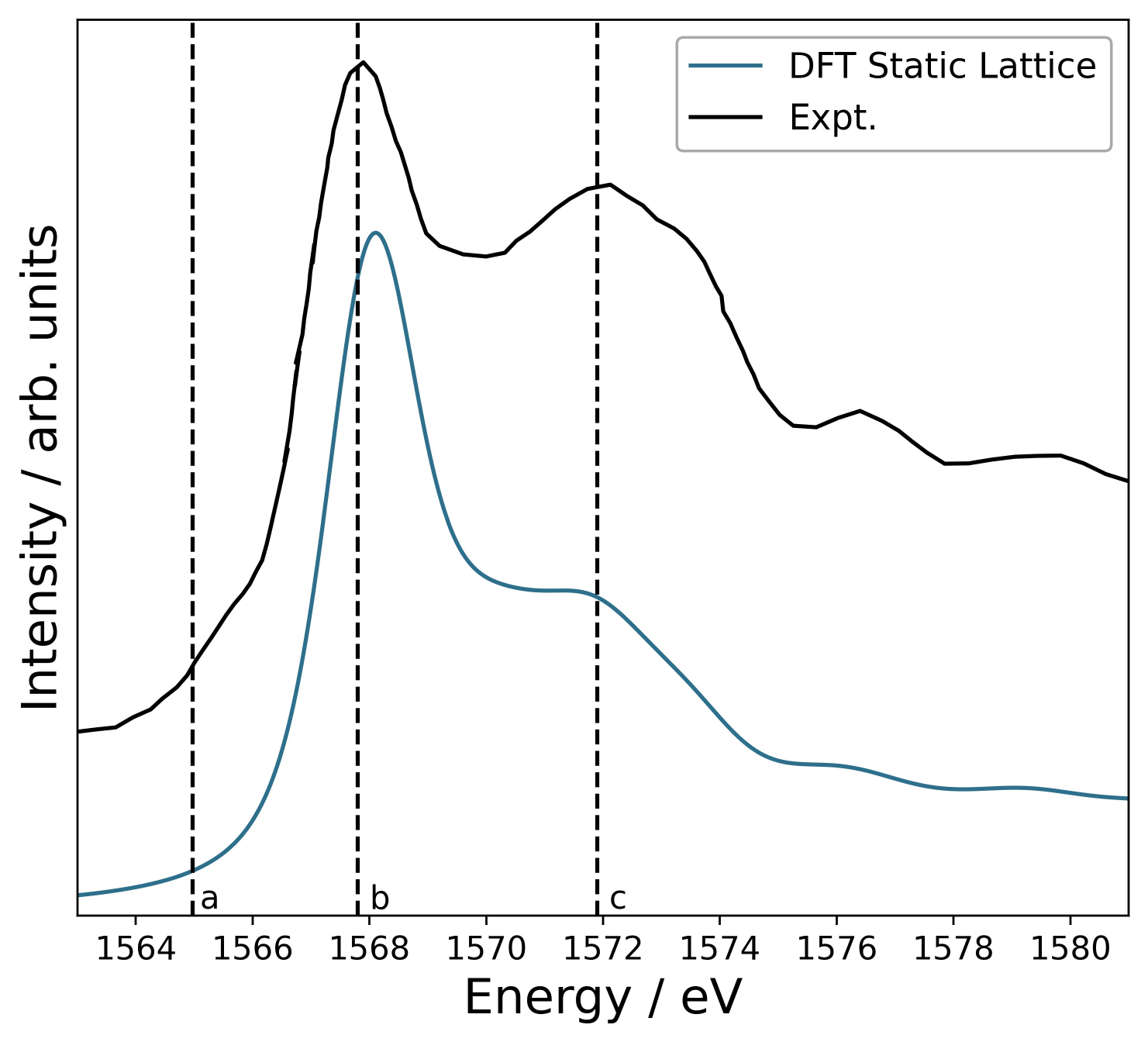}
    \caption{}
    \end{subfigure}
    \begin{subfigure}{.5\textwidth}
    \includegraphics[width=\linewidth]{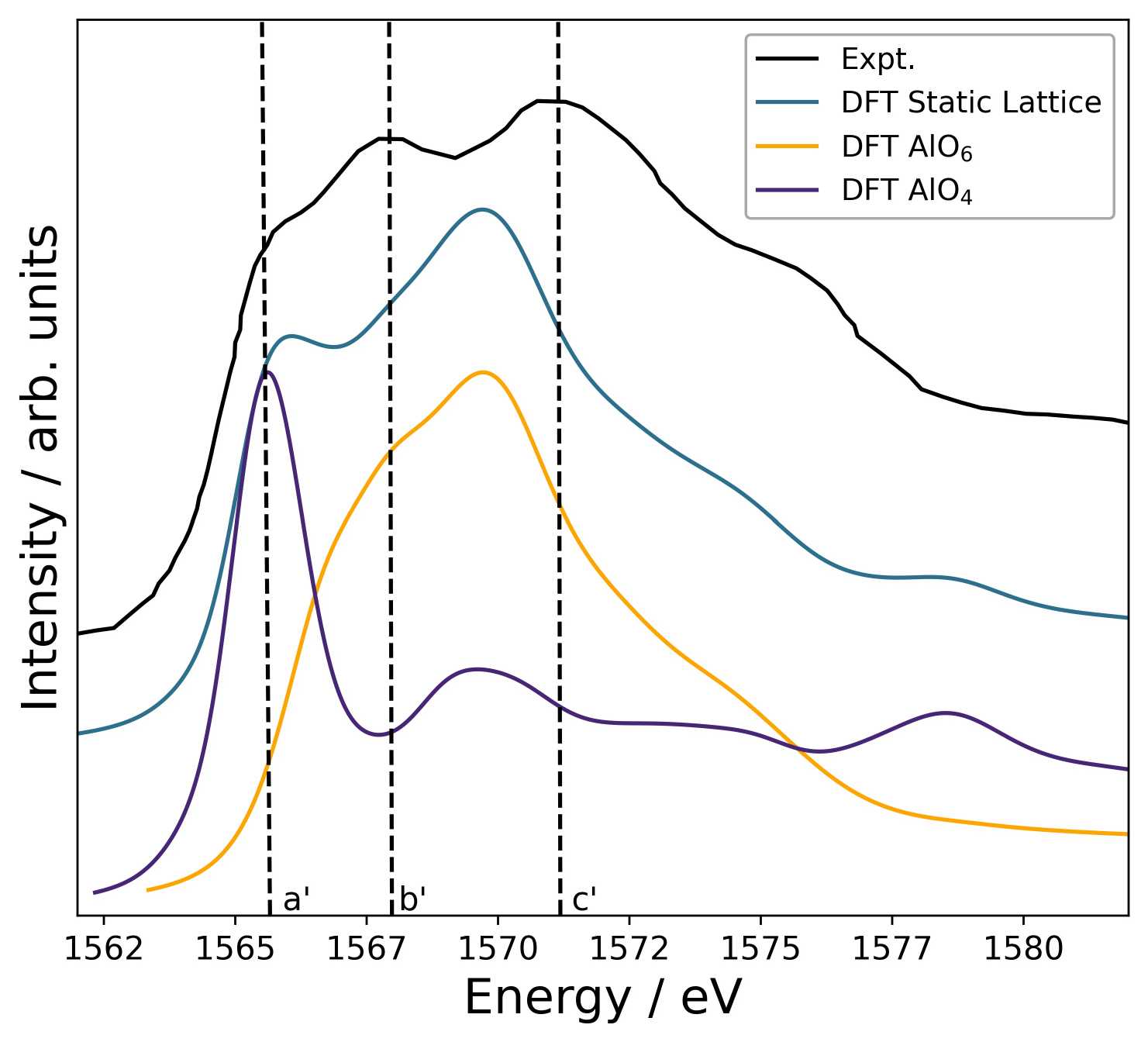}
    \caption{}
    \end{subfigure}
    \caption{(a) $\alpha$-Al$_2$O$_3$ XAS at $0$\,K with a core-hole placed on one Al atom in a 2x2x2 supercell containing 168 atoms. (b) Static lattice XAS for $\gamma$-Al$_2$O$_3$ calculated as the weighted average over two spectra from core-holes placed on AlO$_4$ and AlO$_6$ sites. The $\gamma$-Al$_2$O$_3$ ground state unit cell used for this calculation is the Pinto \textit{et al.} monoclinic spinel-based model \cite{PintoGammaAl2O3} which contains a ratio of AlO$_4$:AlO$_6$ of 6:10, in line with the Smr\v{c}ok cubic spinel model with partial occupancy. Experimental spectrum adapted with permission from Altman \textit{et al}. \textit{Inorg. Chem.} 2017, 56, 10, 5710–5719 Copyright 2017 American Chemical Society.}
    \label{fig:gamma-static-xas}
\end{figure*}

The DFT calculated static lattice level XAS spectrum for both $\alpha$- and $\gamma$-Al$_2$O$_3$ are calculated and shown in Figure \ref{fig:gamma-static-xas}. In the case of $\alpha$-Al$_2$O$_3$, the pre-edge peak at \textbf{a} is not present in the static lattice spectrum, however peaks \textbf{b} and \textbf{c} are both in line with experiment. This spectrum is calculated with a core-hole placed on one Al site within the 2$\times$2$\times$2 supercell of $\alpha$-Al$_2$O$_3$, as there is only one symmetric AlO$_6$ site within the unit cell. On the other hand, $\gamma$-Al$_2$O$_3$ contains one AlO$_4$ site and one AlO$_6$ site (as shown in Figure \ref{fig:structures}b) and therefore two core-hole calculations were performed in the $\gamma$-Al$_2$O$_3$ case. These are shown in Figure \ref{fig:gamma-static-xas}b, in which the total spectrum is calculated as the sum of both the AlO$_4$ and AlO$_6$ sites, weighted by their abundance of 6:10 respectively. The \textbf{a$'$} peak in this spectrum is the only peak resolved at the static lattice level of theory.

\subsection{Electronic DOS and corresponding spectra}

\begin{figure*}[h!]
    \centering
    \includegraphics[width=0.5\textwidth]{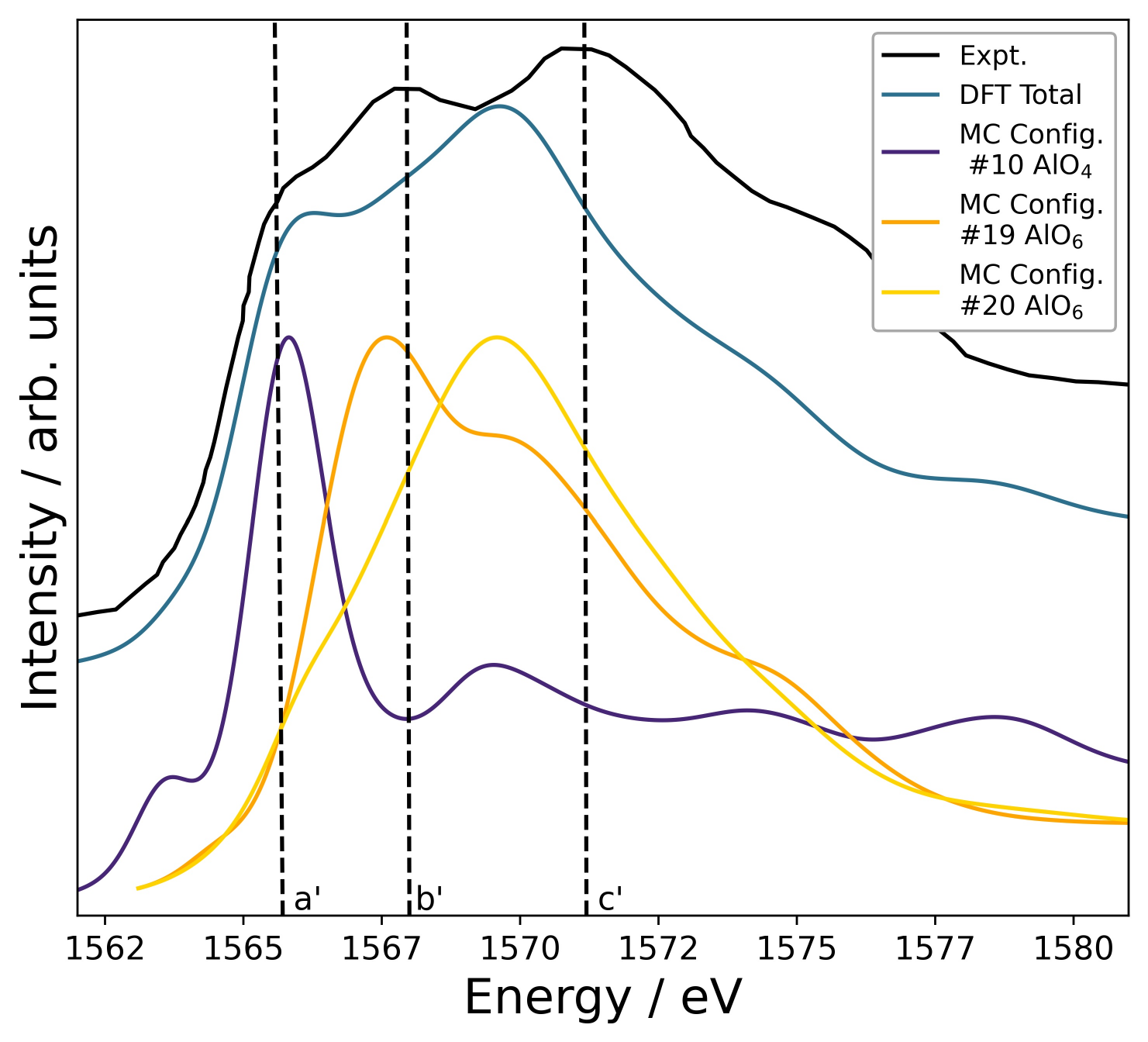}
    \caption{Three configurations from the Monte Carlo sampled finite temperature XAS for $\gamma$ alumina used to calculate the DOS shown in the main text. The configurations are numbered according to their corresponding numbers assigned in the MC sampling. The purple spectrum shows snapshot 10, with a core-hole placed on the AlO$_4$ environment. The orange and yellow lines show configurations 19 and 20 respectively, with a core-hole placed on the AlO$_6$ environment. The DFT Total spectrum is shown for reference, and is the weighted average over all 30 configurations from the MC sampling. }
    \label{fig:supp-gamma-10-19-20}
\end{figure*}

\begin{figure*}[h!]
    \centering
    \includegraphics[width=0.6\textwidth]{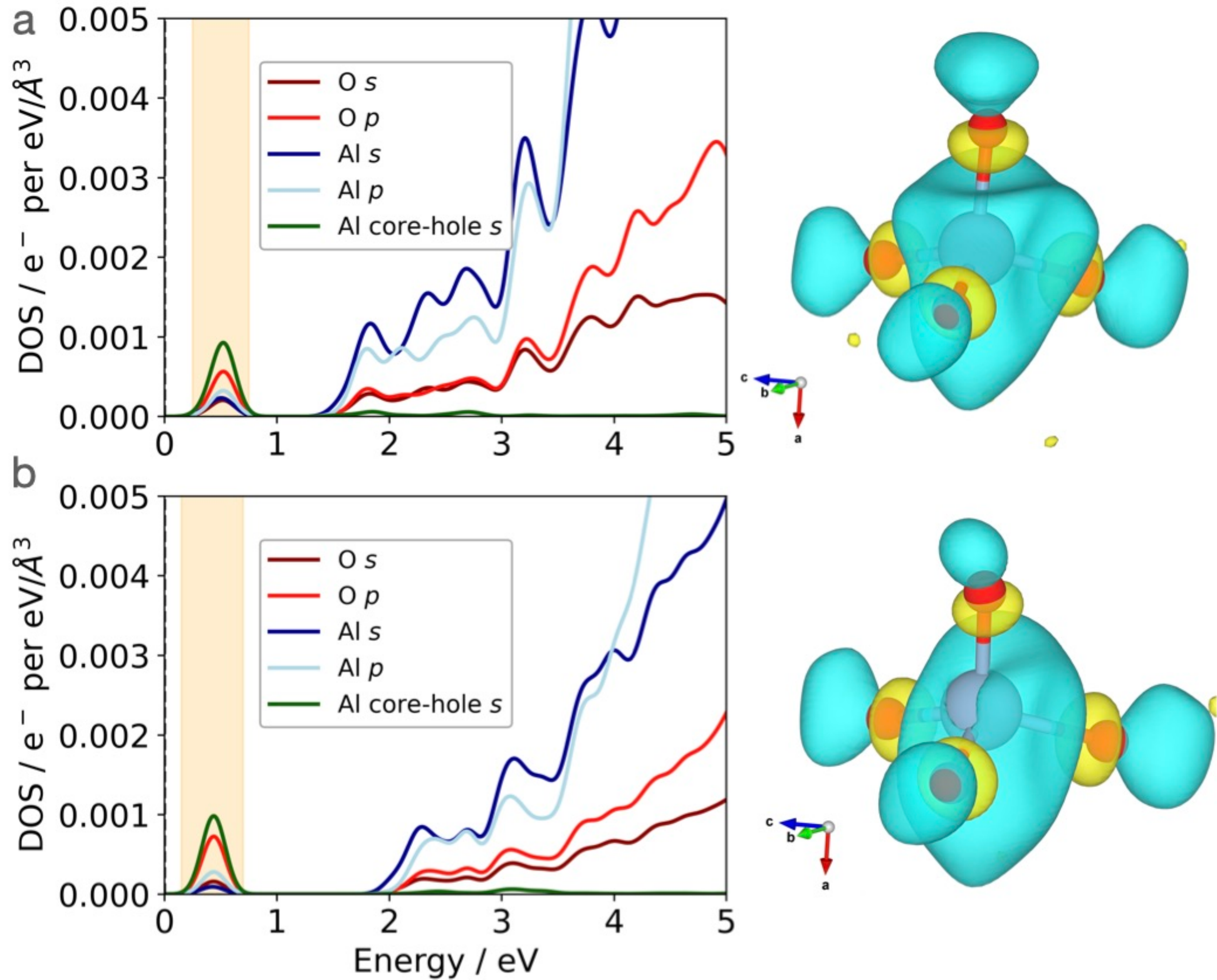}
    \caption{(a) Electronic DOS for the static lattice $\gamma$-Al$_2$O$_3$ phase with a core-hole placed on the AlO$_4$ site. This phase shows a symmetric distribution of the p-orbitals on the four surrounding oxygen atoms from the tetrahedral Al atom (as shown in the right panel) from the lowest unoccupied band in the DOS at $0.9$\,eV above the Fermi level. The Al atom is shown at the center of this image in blue, surrounded by red O atoms, and the positive and negative electron density in 3D are shown. (b) Electronic DOS for configuration 10 from MC sampling (see Figure \ref{fig:supp-gamma-10-19-20} for the XAS for this phase) shows that the bandgap decreases in the distorted case, and the shape of the electronic density surrounding the central Al atom is distorted, with a lower electronic density on the O atom in the -c axis direction when compared to the static lattice case. Both the static lattice and configuration from MC sampling show a maximum XAS peak at \textbf{a$'$} (1565.7\,eV) and the configuration from MC sampling has a pre-edge at $1563$\,eV, not seen in the static lattice case.}
    \label{fig:supp-gamma-alo4-dos}
\end{figure*}

The electronic DOS shown in Figure \ref{fig:gamma-dos} corresponds to the DOS for the static lattice unit cell with a core-hole on an AlO$_6$ atom, and two configurations from the MC sampling at 300\,K. These are MC configurations numbered 19, and 20, and their corresponding absorption spectra are shown in Figure \ref{fig:supp-gamma-10-19-20}. All MC sampled configurations with a core-hole on the AlO$_4$ atom have a main peak at \textbf{a$'$}, however a select set of configurations additionally contain a pre-edge peak. The corresponding XAS spectrum for MC configuration number 10 with a core-hole placed on the AlO$_4$ atom is shown in Figure \ref{fig:supp-gamma-10-19-20}. Comparing the DOS for the static lattice configuration with configuration number 10 (see Figure \ref{fig:supp-gamma-alo4-dos}) shows that there is a distortion of the electron density surrounding the core-hole atom, which results in a slightly higher mixing between the O $p$ states and the Al $s$ states, resulting in a pre-edge peak for this configuration. 

\end{document}